\documentstyle[aps,preprint]{revtex}
\begin{document}

\title{Restoration of Isospin Symmetry in Highly Excited Nuclei}

\author{H. Sagawa$^1$, P.F. Bortignon$^2$ 
	and G. Col\`o$^2$}

\address{$^1$ Center for Mathematical Sciences, University of Aizu, 
	 Aizu-Wakamatsu, Fukushima 965, Japan}

\address{$^2$ Dipartimento di Fisica, Universit\`a degli Studi and INFN, 
	 Sezione di Milano, Via Celoria 16, 20133 Milano, Italy}

\date{\today}

\maketitle

\smallskip

\begin{abstract}
Explicit relations between the isospin mixing probability, the
spreading width $\Gamma_{IAS}^{\downarrow}$ of 
the Isobaric Analog State (IAS) and the 
statistical decay width $\Gamma_c$ of the compound nucleus 
at finite excitation energy, are derived 
by using the Feshbach projection 
formalism. The temperature dependence of the isospin mixing
probability is discussed quantitatively 
for the first time by using the values
of $\Gamma_{IAS}^{\downarrow}$ 
and of $\Gamma_c$ calculated by means of microscopic models.
It is shown that the mixing probability remains essentially constant 
up to a temperature of the order of 1 MeV and then decreases
to about 1/4 of its zero temperature value, at higher temperature than 
$\approx$ 3 MeV, due to the short decay 
time of the compound system.\\
\end{abstract}

\vspace{3.0cm}

PACS and Keywords: 21.10.Hw, 25.70.Gh, 24.30.Cz; isotopic spin, symmetry
restoration, compound nucleus. 

\newpage

Isospin symmetry is the first dynamical symmetry in nuclear physics
proposed by W. Heisenberg in 1932~\cite{Hei32} and has played a very
important role in the study of the structure of atomic nuclei for many
years. One of the highlight of the isospin symmetry is the discovery 
of the isobaric analog state (IAS) of the parent ground state in 
the daughter nucleus by means of charge exchange 
reactions~\cite{IAS_exp61}. Recently, problems related to 
the isospin symmetry and its breaking have been discussed
with new emphasis because new experiments in nuclei 
near the drip lines have become possible. These radioactive beams 
facilities can provide access to regions of large isospin 
deformations~\cite{general}. Several theoretical 
predictions have been made for the isospin mixing and large 
mixing probabilities in heavier nuclei near the proton drip line 
are found, up to (5-6)\% in $^{100}$Sn~\cite{ISO_th}. 
Possible effects of these 
large isospin impurity are at present under consideration, both by 
experimentalists and theoreticians, in connection 
with the width of the IAS and the Fermi $\beta^-$ decay of N$\cong$Z nuclei.  

A restoration of the isospin symmetry in highly excited compound 
nuclei was pointed out by H. Morinaga~\cite{Mori55} and 
D.H. Wilkinson~\cite{Wil56} already 40 
years ago. This problem has been studied experimentally by 
measuring the cross sections for the de-excitation of the 
giant dipole resonances 
(GDR) in thermal, self-conjugate 
nuclei. The experimental data show the sharp decrease of the mixing 
amplitude at high excitation energies above $E^*$=20 MeV in
$^{28}$Si and $^{60}$Zn~\cite{Sno93} as well as 
in $^{40}$Ca~\cite{Cin95}. 
The restoration of the isospin symmetry was theoretically discussed by using 
a simple model in Ref.~\cite{Zel97}.

In this letter, we would like to address the problem of the 
restoration of the isospin symmetry 
by using a microscopic model, which was recently proposed and used at 
zero temperature in Ref.~\cite{SSC96}. This model is based on the   
Feshbach projection method and gives an explicit 
relation between the spreading width of the IAS and the isospin mixing 
probability. In Ref.~\cite{SSC96} this relation was employed successfully 
to calculate, from the values of the isospin mixing, the spreading widths 
of the IAS in Sn- and I-isotopes and in $^{208}$Bi. In 
the present work, we will extend the microscopic model to the case of 
excited compound nuclei, and study quantitatively the temperature dependence
of the isospin mixing through its relation with the spreading width of 
the IAS.   

The isospin mixing amplitude is written in perturbation theory as~\cite{Aue83} 
\begin{equation}
 \alpha^{T_0 + 1} =\frac{1}{E^{T_0}-E_{M}^{T_0+1}} 
 \langle M; T=T_0 +1,T_z = T_0  \mid H_1 \mid T=T_0 ,T_z = T_0  \rangle,
\label{isomix}\end{equation}
where $H_1 $ includes the isospin non-conserving interactions, i.e., the 
isovector part of the Coulomb 
interaction and the charge symmetry breaking as well as charge independence 
breaking 
interactions. The state with isospin $T_0$ is the mother state (either 
the ground state or the IAS), while the states with isospin $T_0 + 1$ give rise to the isospin impurity 
in the mother state by coupling to it.  In the cases of medium-heavy and   
heavy nuclei, the isovector giant monopole resonances (IVGMR) 
give the major contribution to the isospin mixing so that we consider 
only the coupling to these states in the following. 

At finite temperature the mother state lies at 
some high excitation energy and its lifetime as well as that of the 
IVGMR built on top of it, are affected by statistical 
emission of particles (mainly neutrons) and $\gamma$-rays. 
The corresponding width is the compound width $\Gamma_c$ and 
the denominator of Eq.~(\ref{isomix}) should contain the sum of the 
imaginary terms  
$i\Gamma_c^{T_0}/2$ and $i\Gamma_c^{T_0+1}/2$~\cite{Landau}. We simply 
denote in the following this sum as $i\tilde \Gamma_c/2$ and 
$\tilde \Gamma_c$ will be 
called ``compound width''. In general, we put a tilde over the widths 
at finite temperature. The compound width will be 
added to the sum $\tilde \Gamma_M$ of the intrinsic escape and spreading 
widths of the $T_0+1$ monopole 
state. We discuss later how all these 
widths are estimated by using microscopic models.    

Let us now study the relation between the isospin mixing amplitude and 
the spreading width of the IAS. We will use the same Feshbach projection 
method of Ref.~\cite{SSC96} by extending it to the case of 
excited compound nuclei.
The whole model space of nuclear configurations is built up with states 
which are eigenstates of the isospin-conserving nuclear Hamiltonian $H_0$ 
and is divided in two subspaces P and Q. The P space contains only
two states, the parent ground state and the IAS which have 
isospin $T_0$. The other nuclear states in the daughter nucleus which have 
isospin $T_0-1$ form the space Q. The spreading width of the IAS 
($\tilde{ \Gamma}_{IAS}^{\downarrow}$) is 
obtained by coupling with doorway states of the space Q 
through the isospin-breaking Hamiltonian as~\cite{SSC96}
\begin{equation}
 \tilde{ \Gamma}_{IAS}^{\downarrow} 
 (E^*) = -2 Im \langle \mbox{IAS} \mid H_{1}
 Q \frac{1}{E^* - H_{QQ}} Q  H_{1} \mid \mbox{IAS} \rangle \\ 
\label{gammadown_gen}\end{equation} 
where $E^*$ is the energy of the IAS in the compound nucleus and 
$H_{QQ}$ is the matrix element of the isospin-conserving Hamiltonian 
$H_0$ between states in Q-space, among which the dominant role is 
played by the three isospin components of the IVGMR in the daughter nucleus. 
At finite temperature, both the 
states in P- and Q-spaces have finite decay times, as discussed above.
In terms of 
$\tilde{\Gamma}_{c}$ and 
$\tilde{\Gamma}_M$, Eq.~(\ref{gammadown_gen}) is rewritten as 
\begin{eqnarray}
 \tilde{\Gamma}_{IAS}^{\downarrow} & = & -2 Im \sum_{q} \frac{\mid \langle 
 \mbox{IAS} \mid H_{1} \mid q  
 \rangle \mid^{2}}{E^* - E_{q} 
 + i \tilde{\Gamma}_{c}(E^* )/2   
 + i \tilde{\Gamma}_{M}(E^* )/2} \nonumber\\ 
 &=& \sum_{q} ( \tilde{\Gamma}_{c}(E^* ) + \tilde{\Gamma}_{M}(E^* ) ) 
 \frac{\mid \langle \mbox{IAS} \mid H_{1} 
 \mid q \rangle \mid^{2}}{(E^* -E_{q})^{2} + 
 (\tilde{\Gamma}_{c}(E^* )/2 + \tilde{\Gamma}_{M}(E^* )/2)^{2}} 
\label{gammadown_ii}\end{eqnarray}
where the $E_q$ are the excitation energies of the 
three different isospin components of the IVGMR in daughter nuclei, which 
have $T'$=$T-1,\ T$ and $T+1$ and $\tilde{\Gamma}_{c}(E^* )$ means, as 
recalled above, the sum of the two compound widths at the excitation energy 
corresponding to the nulear temperature, and at that energy plus the 
IAS energy $E^*$.   
The isospin structure 
of the IVGMR can be treated properly by the Feshbach projection method as it 
has been done in Ref.~\cite{SSC96}. The right hand side of 
Eq.~(\ref{gammadown_ii}) results as a product of three factors, by assuming 
that for $\tilde{\Gamma}_{c}$ and $\tilde{\Gamma}_{M}$ average values over 
$q$ can be taken, and precisely 
\begin{equation}
 \tilde{\Gamma}_{IAS}^{\downarrow} = ( \tilde{\Gamma}_{c}(E^* ) + 
 \tilde{\Gamma}_{M}(E^* )) 
 (\alpha ^{T_0 +1})^2
 (T_0 +1)F(T_0 ), 
\label{gamma1}\end{equation}
where the isospin factor $F(T)$ is given by
\begin{equation}
 F(T)=\frac{1}{T} \{ (\frac{2T-1}{2T+1}) 
 \frac{1}{(\Delta E_{M}^{T-1})^{2}}  
 + \frac{(T-1)^{2}}{T+1} \frac{1}{(\Delta E_{M}^{T})^{2}} 
 + \frac{4T^{2}}{(2T+1)(T+1)} \frac{1}{(\Delta E_{M}^{T+1})^{2}} \}
 (\Delta E_{M\pi})^{2}.
\end{equation}
In this last Equation, $\Delta E_{M}^{T'} = E_{M}^{T'} - E^*$ is 
the energy difference between the three isospin components of the 
IVGMR in the daughter nucleus and the energy of the IAS whereas 
${\Delta E_{M\pi}}=E_{M}^{T+1}-{E_{\pi}}$ is the energy difference 
between the $T+1$ component of the IVGMR in the mother nucleus and 
the mother state $|\pi\rangle$. This last difference enters because the 
isospin mixing amplitude 
$\alpha ^{T_0 +1} $ is obtained by Eq.~(\ref{isomix}). In 
Ref.~\cite{SSC96} the isospin splitting was taken care of by using the Lane 
potential and we write a similar expression for nuclei at finite 
temperature,  
\begin{equation}
 {\Delta E}^{T'}_M = E^* (IVGMR) + \frac{V_1
 }{A}{\bf t \cdot T_c}, 
\end{equation}
where $E^* (IVGMR)$ is the average energy of the IVGMR in the compound 
nucleus and ${\bf t} ({\bf T_c })$ is the isospin of the IVGMR (the daughter 
nucleus).  The value $E^*(IVGMR)$ is taken from the systematics 
in nuclei at zero temperature as
\begin{equation}
 E^* (IVGMR) = 170\ A^{-1/3}\ {\rm MeV}. 
\end{equation} 
The value of the parameter of the Lane potential commonly used is 
$V_{1}=120$ MeV. In general, the Brink-Axel 
hypothesis suggests that giant resonances built on excited states lie 
at the same energy as the corresponding resonances built on the ground
state. In particular, in the case of the isovector giant dipole 
resonance it has been experimentally observed that the excitation energy 
does not vary appreciably with increasing temperature~\cite{Gaa92}. 
Thus, it is expected that the energy of the IVGMR also does not vary with
increasing temperature. 

The isospin geometrical factor $F(T_0)$ is certainly important to predict 
a proper 
isotope dependence of the spreading width 
$\tilde{\Gamma}_{IAS}^{\downarrow}$ of several nuclei  
in medium-mass region. For our present purpose, however, we may 
be allowed to 
discard the isospin dependence of the IVGMR energies. 
Then, Eq.~(\ref{gamma1}) becomes
\begin{equation}
 \tilde{\Gamma}_{IAS}^{\downarrow} \cong ( \tilde{\Gamma}_{c}(E^* ) + 
 \tilde{\Gamma}_{M}(E^* ))
 (\alpha ^{T_0 +1})^2 (T_0 +1)
\end{equation} 
so that the isospin mixing probability is given by 
\begin{equation}
 (\alpha ^{T_0 +1})^2=\frac{1}{T_0 +1}
 \frac{\tilde{\Gamma}_{IAS}^{\downarrow}(E^*)}{\tilde{\Gamma}_{c}(E^* ) + 
 \tilde{\Gamma}_{M}(E^* )}. 
\label{final}\end{equation}
In Eq.~(\ref{final}), $\tilde {\Gamma}_{c}$ is expected to rise significantly 
with temperature. On the other 
hand, the values of the spreading width $\tilde{\Gamma}_{IAS}^{\downarrow}$ 
and of the total width $\tilde{\Gamma}_{M}$ are not expected to be very 
much temperature dependent, because of the following reasons. 
The temperature dependence of the intrinsic 
escape widths $\tilde{\Gamma}^\uparrow$ of giant resonances 
is discussed in Ref.~\cite{Sag84} and 
is found to be small. For the intrinsic spreading widths 
$\tilde{\Gamma}^\downarrow$, the authors of Ref.~\cite{Don96} predict 
a very mild temperature dependence. 
Consequently, from Eq.~(\ref{final}) we should expect 
at high temperature a strong decrease of the isospin 
mixing, or equivalently a restoration of the isospin 
symmetry. Eq.~(\ref{final}) certainly justifies the 
conjecture which was raised by Wilkinson~\cite{Wil56} to describe 
the isospin mixing problem: if the compound 
nucleus decays on a time scale which is shorter than 
the time needed for a well-defined isospin state to mix with  
states with different isospin, then the isospin symmetry is partially 
or totally restored. Consequently, the
restoration of the isospin symmetry can be understood on the ground of
simple kinetic arguments (as pointed out also in Ref.~\cite{Zel97}). The
same kinetic arguments are used in Ref.~\cite{Bor91}, where an equation with
the same physical content as our Eq.~(\ref{final}) is obtained within the
context of the study of the disappearance of the collective giant resonance
strength at high excitation energy.

The isospin mixing in compound nuclei was also discussed extensively in 
Ref.~\cite{HRW}. The authors derived a formula for the isospin mixing 
probability $\alpha ^2 $ in self-conjugate nuclei as
\begin{equation}
 (\alpha ^{T_0 +1})^2 = 
 \frac{ \Gamma_{2}^{\downarrow}}{ \Gamma_{2}^{\uparrow}}, 
\label{Harney}\end{equation}
where $\Gamma_{2}^{\downarrow}$ and $\Gamma_{2}^{\uparrow}$ 
are the mixing and total decay width of the states in the channel 2 
which is equivalent to the Q-space of our model. Although they do not specify 
the spin and parity of the states in channel 2, the 
formula~(\ref{Harney}) looks similar to our Eq.~(\ref{final}) except 
for the isospin 
factor which is necessary for non self-conjugate nuclei.     
It should be noticed that the two models give almost the same final result, 
although the formalisms by us and by the authors of~\cite{HRW} look at first 
sight very different. 

To discuss quantitatively the dependence of the isospin mixing on the 
excitation energy of the compound nucleus, in the example of  
$^{208}$Pb, we have to model the 
temperature dependence of all quantities which appear in
Eq.~(\ref{final}): the intrinsic widths of the IVGMR, 
$\tilde\Gamma^\uparrow_M$ and $\tilde\Gamma^\downarrow_M$, and 
the intrinsic spreading width of the IAS $\tilde\Gamma^\downarrow_{IAS}$, 
as well as the compound width $\tilde\Gamma_c$. 
First, the intrinsic spreading widths 
$\tilde\Gamma^\downarrow_M$ and $\tilde\Gamma^\downarrow_{IAS}$ are 
parametrized by means of a 
linear temperature dependence according to the findings 
of~\cite{Don96}, 
\begin{equation}
 \tilde{\Gamma}^\downarrow = \Gamma^\downarrow (T=0)\cdot (1+cT), 
\label{tempdep}\end{equation}
where $T$ is here the nuclear temperature in units of MeV and $c$ is 
a constant which should be determined using the results of a microscopic 
calculation. In general, giant resonances are described microscopically 
as a coherent superposition of one particle-one hole (1p-1h) excitations. 
The main cause of their spreading width stems from the coupling to 
collective low-lying surface vibrations. In 
Ref.~\cite{Don96}, the spreading width of  
the GDR in $^{120}$Sn has been studied and the slope parameter $c$ has 
been estimated to be about 0.025 or less, by 
fitting the results of microscopic calculations which take into account  
the coupling of the giant resonance with states composed of a 1p-1h pair plus a 
low-lying vibration~\cite{Bor86}. These doorway states to which 
the giant resonance is coupled directly, are 
not considered to be sharp but distributed over an energy 
interval of the order of $\Delta$ with a Lorentzian shape. In this 
spirit, the width $\Delta$ is supposed to take care of the coupling of 
the doorway states with the compound nucleus states. This coupling has been 
shown, on the basis of rather general hypothesis, to be essentially constant 
as a function of temperature~\cite{PRL}. On the other hand, we may let  
$\Delta$ grow when increasing $T$ to simulate the fact that the 
doorway states are more fragmented since the low-lying states become  
less collective~\cite{Bor86}. It is found~\cite{Col_tbp} that the change of
the spreading width of the IAS is at most 20\% when $\Delta$ is varied from 
50 keV to 500 keV, and the spreading width saturates for larger values of
$\Delta$. If this 20\% change of the spreading width 
$\tilde{\Gamma}^\downarrow_{IAS}$ is supposed to occur within 
the temperature range 
between 0 and 4-5 MeV, the slope parameter $c$ is $c$ = 
0.04-0.05, which is not much different from the value mentioned above, 
$c$ = 0.025. Consequently, we take the linear dependence~(\ref{tempdep}) 
with $c$ = 0.025 for $\tilde\Gamma^\downarrow_M$ and 
$\tilde\Gamma^\downarrow_{IAS}$. Then, we show below that our conclusions 
are not markedly changed even if we increase by a factor 2 our value of $c$. 

At zero temperature, the spreading width of the 
IVGMR $\Gamma^\downarrow_M(T=0)$ in $^{208}$Bi, at the excitation 
energy of the IAS, is found to be of the order of 250 keV~\footnote{This 
calculation was already reported in Ref.~\cite{SSC96} where the total width 
of the IVGMR in $^{208}$Bi at the IAS energy was estimated to be about 450 
keV: about a half of the total value, that is, 250 keV, corresponds to the 
spreading width whereas 200 keV corresponds to the escape width.}, while 
that of the IAS $\Gamma^\downarrow_{IAS}(T=0)$ is evaluated 
to be about 100 keV~\cite{Col_tbp} including the contribution of the 
charge-symmetry breaking and charge-independence breaking 
interactions~\cite{CSB_CIB}. This value $\Gamma^\downarrow_{IAS}
(T=0)$ = 100 keV is in good agreement with the experimental 
value~\cite{HRW,IAS_exp}. Finally, $\tilde{\Gamma}^\uparrow_M(T)$ is taken 
as a constant according to the findings of Ref.~\cite{Sag84} and fixed at 
200 keV~\footnote{See the previous note.}.  

The compound widths $\tilde\Gamma_c$ are dominated by the most probable decay
process of the compound nucleus which is usually neutron evaporation. We may
discuss the neutron evaporation in simple terms, as it is done in
Ref.~\cite{Bri90} where a Weisskopf formula is used for the
temperature dependence of $\tilde\Gamma_n$,
\begin{equation}
 \tilde\Gamma_n = {2mR^2\over\pi\hbar^2}T^2 e^{-B_n/T},  
\label{Weisskopf}\end{equation}
with a radius $R$ = 7.1 fm and a neutron separation energy $B_n$ = 7.4 MeV
in the case of $^{208}$Pb. It is known that the neutron widths obtained 
through this simple formula may be somewhat different from more realistic 
neutron widths, essentially due to the schematic
assumption of a geometric absorbtion cross section for neutrons on
the nucleus. In order to take into account the microscopic transmission 
coefficient, we have also estimated the compound widths 
by means of the statistical computer code CASCADE~\cite{Pul77}. This 
code takes into account also proton and $\alpha$ emission and needs as an 
input the
excitation energy $E^*$ of the compound system which can be related to the
nuclear temperature by the well-known relation $E^* = aT^2$ for which we have
taken $a$ =10 MeV$^{-1}$. The results for the isospin mixing (\ref{final}) 
are not 
so markedly sensitive to whether we use compound widths produced by CASCADE 
or by the simple formula (\ref{Weisskopf}). Therefore, for simplicity we 
will use the values of $\tilde\Gamma_c\equiv\tilde\Gamma_n$ obtained by 
using Eq.~(\ref{Weisskopf}). 
  
We present our results for the isospin mixing probability of $^{208}$Pb as 
a function of temperature in Fig.~1. All curves start, at zero temperature, 
from the value $\alpha^2$ = 0.1, corresponding to a value of the spreading 
width of the IAS, $\Gamma^\downarrow_{IAS}$ = 100 keV. The solid line 
corresponds to the calculation described above with the slope 
parameter $c$ = 0.025. The dashed line corresponds to 
choosing the parameter $c$ equal to 0.05 both for the 
IVGMR and for the IAS. These two results show that our general 
conclusions are not too much affected by the choice of $c$ and that the 
isospin restoration starts to take place at a temperature of  
about 1.5 MeV. At 3 MeV temperature, the value of the isospin mixing 
probability has decreased to one fourth of 
the value at zero temperature due to the rapid increase 
of the compound width appearing in the denominator of Eq.~(\ref{final}). 

Experimentally, as mentioned above, the temperature dependence of the 
isospin mixing in the decay of the GDR was studied in the light nuclei 
$^{28}$Si and $^{60}$Zn~\cite{Sno93}. A 
similar substantial decrease of the isospin mixing was found
above $T$=1 MeV although the magnitude of the mixing at low
temperature is much larger than in the present case because 
of the coupling to the shell model $T_>$ states, which is more important  
than the coupling to the IVGMR in light nuclei~\cite{Mekjian}. 

In summary, we have studied the isospin mixing in compound nuclei at finite 
temperature with a microscopic model based on the Feshbach projection method.  
We are able to derive a relation which connects three physical quantities, 
that is, the isospin mixing amplitude, the spreading 
width of the IAS and the statistical decay width of the compound system. 
In this way, the isospin mixing probability in $^{208}$Pb as a function of 
the temperature is studied quantitatively for the first time by using the 
results of microscopic calculations. It is shown that until relatively low 
excitation energies, corresponding to a temperature of the order of 1 MeV, 
the mixing probability remains constant. At higher excitation energies,  
this mixing probability decreases and around 3 MeV it is reduced by 
about a factor 4 because of the rapid increase of the compound width.   
Thus, we have verified, by using microscopic 
ingredients, the validity of the conjecture formulated by Wilkinson 40 years  
ago, of the restoration of the isospin symmetry in highly excited systems. 
In connection with the present work, it would be of extreme importance to 
obtain experimental information about the width of the IAS at finite 
temperatures.

\vspace{1.5cm} 

H.S. likes to acknowledge the nice hospitality of the Universit\`a degli 
Studi in Milano where part of the work has been done, as well as financial 
support from INFN. P.F.B. and H.S. like to acknowledge the hospitality 
of the Institute for Nuclear Theory of the University of Washington of 
Seattle, where some original ideas of the present work were discussed and 
stimulating discussions with K.A. Snover took place.

\newpage

% Figure caption

Fig.1\\
The temperature dependence of the isospin mixing probability $\alpha^2$ in 
$^{208}$Bi. This probability is calclulated by using Eq.~(\ref{final}), 
and assuming a linear temperature dependence for 
$\tilde\Gamma^\downarrow_M$ and $\tilde\Gamma^\uparrow_{IAS}$ according to 
Eq.~(\ref{tempdep}). The slope parameters $c$ adopted in~(\ref{tempdep}) are 
0.025, and 0.05 for the solid and dashed curves, respectively. 
The value of $\tilde\Gamma_c$ is evaluated through Eq.~(\ref{Weisskopf}). 

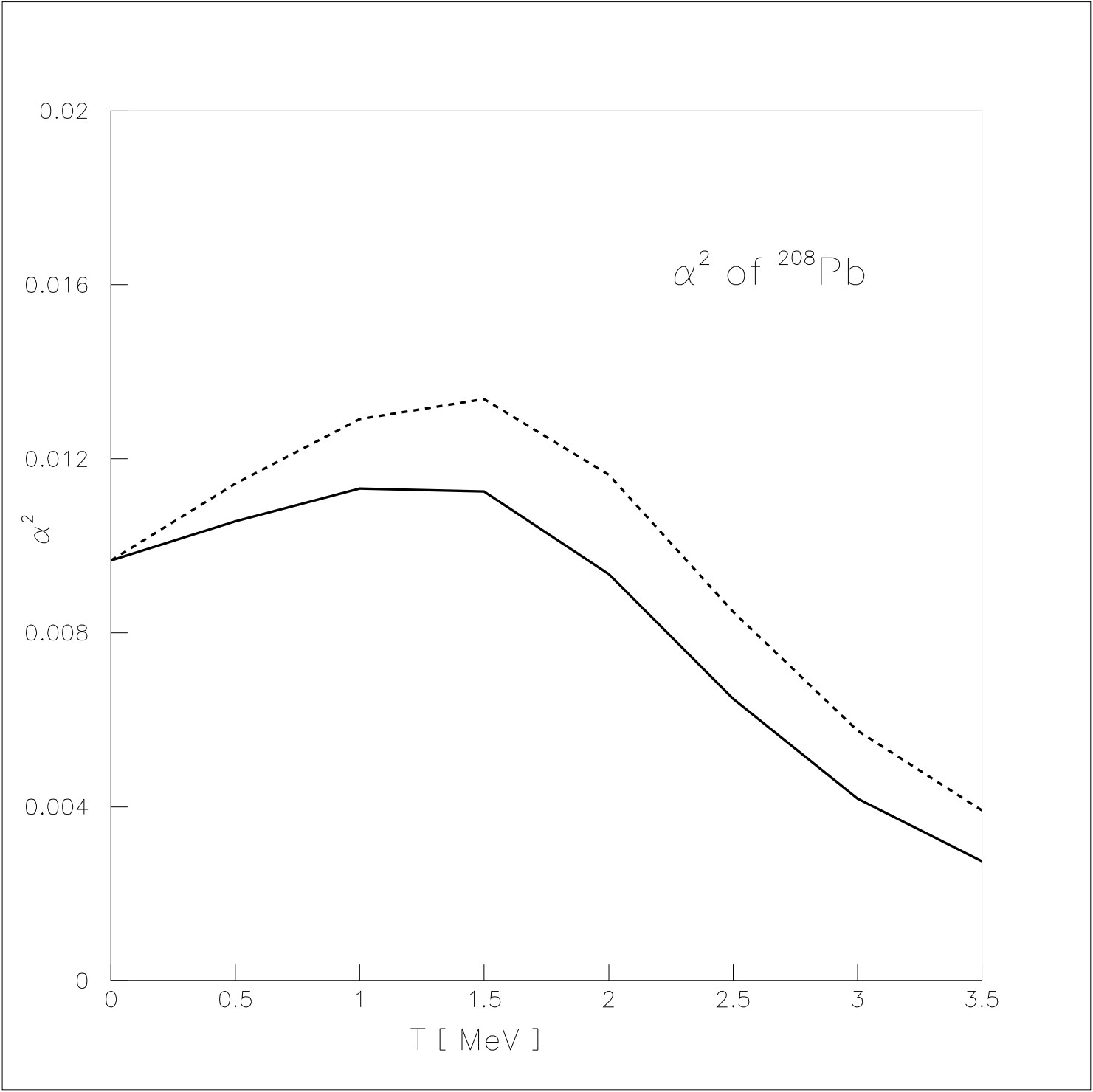
\end{document}